\documentclass[aps,prc,superscriptaddress,showpacs]{revtex4}
\usepackage[dvips]{graphicx}
\newcommand{\psla}{\mbox{\ooalign{\hfil/\hfil\crcr$p$}}}
\newcommand{\ksla}{\mbox{\ooalign{\hfil/\hfil\crcr$k$}}}

\newcommand{\eq}{\label}

\newcommand{\vp}{\mbox{\boldmath $p$}}
\newcommand{\vq}{\mbox{\boldmath $q$}}
\newcommand{\vk}{\mbox{\boldmath $k$}}

\newcommand{\Nbar}{\mbox{${\bar N}$}}
\newcommand{\NNbar}{$N{\bar N}$}
\newcommand{\sNNbar}{N{\bar N}}

\newcommand{\Pitild}{\mbox{${\tilde{\Pi}}$}}

\newcommand{\Stil}{\mbox{${\tilde{S}}$}}
\newcommand{\vPi}{\mbox{${\boldmath \Pi}$}}

\newcommand{\epsi}{\mbox{$\varepsilon$}}

\begin{document}

\title{\bf  
Energy-Dependence of the Nucleon Self-Energies in Off-Mass-Shell Energy Region
and 
the Gamow-Teller Sum-Rule  in the Relativistic Hartree-Fock Approach}

\author{Tomoyuki~Maruyama}
\affiliation{College of Bioresource Sciences,
Nihon University, Fujisawa, Kanagawa 252-8510, Japan}
\affiliation{Advanced Science Research Center, 
Japan Atomic Energy Agency, Tokai, Ibaraki 319-11, Japan }

\begin{abstract}
The relativistic Hartree approximation predicts a deep attractive
 potential for antinucleon, which largely reduce the threshold energies 
of the nucleon-antinucleon (\NNbar) production.
This effect has played an important role to explain the quenching of 
the Gamow-Teller (GT) strength because the quenched strength 
in the particle-hole 
excitation is partially taken by the nucleon-antinucleon production. 
On the other hand antiproton experiments do not reveal deep attractive
 potential for antinucleon.
In this paper we study energy-dependence of the nucleon self-energies 
in the relativistic Hartree-Fock (RHF) approximation in off-mass-shell states.
Then we demonstrate that the antinucleon appearing in low energy 
observables is in the off-mass-shell energy region
 and that  its properties are quite different from that at the on-mass-shell state.
Furthermore we show that the quenched amount of the GT strength does not
shift only to the \NNbar~production but also to the meson production
through the imaginary part of the nucleon self-energy in the RHF approximation. 
\end{abstract}


\maketitle

\vfil
\eject

\newpage

\section{Introduction}

The past decades have seen many successes in the relativistic 
mean-field (RMF) approach of the nuclear many-body problem.
The relativistic framework has big advantages in several 
aspects \cite{Serot}: a useful Dirac phenomenology for the 
description of nucleon-nucleus scattering \cite{Hama,Tjon}, 
the natural incorporation of the spin-orbit force \cite{Serot} 
and the saturation properties in the microscopic treatment with 
the Dirac Brueckner Hartree-Fock (DBHF) approach \cite{DBHF}. 
Furthermore this approach can explain very nicely the structure 
of such extreme nuclei as for neutron-rich nuclei
\cite{hirata}.

These results conclude that there are large attractive scalar 
and repulsive vector-fields, and that the nucleon effective mass
becomes very small in the medium.
These deep scalar and vector fields predict 
very strong attractive potential of antinucleon (\Nbar)
and large suppression of the in-medium threshold energy of
the  nucleon-antinucleon (\NNbar) pair creation  
 because the vector fields is changed to be attractive for \Nbar.   

Kurasawa and Suzuki \cite{ks1,ks4} showed that 
the largely reduced threshold energy of \NNbar~ production in medium 
play a very important role to explain the suppression of the Coulomb sum-rule
in the quasielastic electron scattering.
Furthermore they showed that the Gamow-Teller (GT) transition strength
 is quenched by the antinucleon degree of freedom \cite{ksg1,ksg2}.
The particle-hole ($ph$) excitation explains 88 \% of
the Ikeda-Fujii-Fujita (IFF) sum-rule, and the other 12 \% strength
is given by the \NNbar-production.
The shift of the strength from the $ph$-excitation sector to the 
\NNbar-production sector increases as the effective mass becomes smaller \cite{ks2}.

In the RMF approach, thus, the largely reduced threshold energy 
play important roles in calculations of several low energy phenomena. 
However experimental observables about the antiproton have not
revealed this deep \Nbar~attractive potential.
For example 
the analysis of  the antiproton (${\bar p}$) + nucleus elastic scattering
has not shown  its deep attractive potential \cite{YuShun}.
Furthermore Teis et al. \cite{Teis} analyzed the subthreshold 
${\bar p}$-production experiments in nucleus-nucleus collisions \cite{GSIpbar} 
and light-ion induced collisions \cite{KEKpbar} and concluded that
the ${\bar p}$ mean-field is attractive but its depth is much smaller
than that predicted in the RH approximation.
Particularly the results in the light-ion induces reactions 
gave important results that the the  cross-section in the deuteron
induced reaction, $\sigma_d$, is fifty times larger than that in the
proton induced reaction, $\sigma_p$ at the initial energy, 
$E_{lab} =3.5$ GeV/u: $\sigma_d/\sigma_p \approx 50$. 
If the ${\bar p}$ potential is as deep as that in the RH approximation, this initial
energy is above the threshold energy, and the ratio must be close to
the nucleon number of the induced particles, 
namely $\sigma_d/ \sigma_p \approx 2$.

Of course there are some ambiguities even in theoretical analysis of
experimental results.
For example, G.Q. Li et al. \cite{Lipbar} gave a different conclusion 
about the analysis of the GSI experiments \cite{GSIpbar} though they have
not shown any analysis about the KEK data \cite{KEKpbar}.  
We cannot definitely determine the ${\bar N}$ potential, but
these experimental results suggests suspicion on the treatment
of the antinucleon in the RMF approach.

In the theoretical aspect only the RH approximation suggests
the deep \Nbar~potential, but one can deny it by extending
the theory, such as the relativistic Hartree-Fock (RHF), 
including the nonlocal parts \cite{soutome}.
In the RHF framework, furthermore, the self-energies of nucleon are momentum-dependent
and have different values between the on-mass-shell state and the off-mass-shell
states \cite{KLW1}.
The antinucleon appearing in calculations of low energy
phenomena must be at an off-mass-shell state, and its property may be
different from that at the on-mass-shell state.
If the antinucleon properties at the off-mass-shell state is different,
moreover, we need to examine the transition strength in the \NNbar-production. 
 
In this subject, it has been believed for a long time that 
the momentum-dependence of the Dirac fields is negligible 
in the low energy region, particularly below the Fermi level.
In fact, only very small momentum-dependence has appeared in 
the RHF calculation \cite{RHF}.
In the high energy region, however, the vector-fields must become very small 
to explain the optical potential of the proton-nucleus elastic scattering 
\cite{Hama,KLW1}, and the transverse flow in the heavy-ion collisions 
\cite{TOMO1}.
In addition this momentum-dependence also play an important role to
explain the quasielastic electron scattering \cite{QELMD}.
Even in low energy region, furthermore, the momentum dependence has been
reported to plays important roles to explain the isoscalar giant
quadrupole resonance \cite{tomoGR}, the spatial convection current
\cite{tomoDC}, nuclear structure \cite{typel} 
ssand the pseudo-spin symmetry \cite{Long}.
In addition the DBHF approach also slightly change a saturation
properties if the momentum dependence is introduced \cite{DBHF2}. 

In this paper we examine the energy-dependence of the Dirac mean-field
in the RHF approximation and show the antinucleon
properties in off-mass-shell energy region.
Furthermore we calculate the GT-strength in the same approach.

In Sec.~II we explain structures of the nucleon propagator and
self-energies in the RHF approach.
In Sec.~III we calculate the energy-dependence of the nucleon self-energies
 and demonstrate that the \NNbar-production energy appearing in
 calculations is energy-dependent and different from that in the actual
 production.
In Sec.~IV we calculate the GT strength and demonstrate that this
strength is contributed from three processes, the $ph$-excitation, the
\NNbar-production and the meson production.
Finally we summarize our work in Sec.~V.

\section{Nucleon Propagator with Momentum-Dependent Self-Energies}

The nucleon propagator in the self-energy $\Sigma$ is
usually given by
\begin{equation}
G^{-1} (p) = \psla - M - \Sigma(p),
\label{prop}
\end{equation}
where $\Sigma(p)$ has a Lorentz scalar part $U_s$ and a Lorentz vector
part $U_{\mu}(p)$ as
\begin{equation}
\Sigma(p) = - U_{s}(p) + \gamma^{\mu} U_{\mu}(p).
\end{equation}
For the future convenience we define the effective mass and the kinetic
momentum as
\begin{eqnarray}
M^{*}(p) & = & M - U_{s}(p) ,
\nonumber \\
\Pi_{\mu}(p) & = & p_{\mu} - U_{\mu}(p) .
\end{eqnarray}
Then the detailed form of the nucleon propagator, $G(p)$, is
represented \cite{KLW1} by
\begin{eqnarray}
G(p) &=& G_F (p) + G_D (p)
\nonumber \\
&=& \frac{ \gamma^\mu \Pi_\mu(p) + M^{*}(p)}{\Pi^2(p) - M^{*2}(p) + i\delta }
+ i \pi n(\vp) \theta (p_0) \left[ \gamma^\mu \Pi_\mu(p) + M^{*}(p) \right]
  \delta [ V(p) ] ,
\label{propd}
\end{eqnarray}
where $n(\vp) = \theta(k_F - |\vp|)$ is the momentum distribution 
with the Fermi momentum, $k_F$, and
\begin{equation}
V(p) \equiv \frac{1}{2}  \left[ {\Pi}^2(p) - M^{*2}(p) \right] .
\end{equation}

Now we will rewrite the above nucleon propagator in the spectral
representation.
The wave-function of nucleon, $\psi (p,s)$, with four-momentum $p$ and spin $s$
is defined as a solution of the following Dirac equation
\begin{equation}
\left[ \gamma^\mu  \Pi_\mu(p) - M^* (p) \right] \psi(p,s) = 0  .
\label{Deq1}
\end{equation}
This Dirac equation is equivalent to the following
characteristic equation:
\begin{equation}
\left[ {\vec \alpha}\cdot {\vec \Pi} (p_0;\vp)  - \beta M^* (p_0;\vp) 
+ U_0  (p_0;\vp) \right] \psi(p,s)
 = \lambda   \psi(p,s) .
\end{equation}
There are two kinds of solutions;
one is so-called a positive energy spinor $u (p,s)$ with 
\begin{equation}
\lambda = e_N (p) \equiv
E_N (p) + U_0 (p) = \sqrt{\Pi_v^2 (p) + M^{*2}(p)} + U_0 (p) ,
\end{equation}
and the other is called a negative energy spinor $v (-p,s)$
with 
\begin{equation}
\lambda =  - e_A(p) \equiv
- E_N (p) + U_0 (p) = - \sqrt{\Pi_v^2(p) + M^{*2}(p)} + U_0 (p) ,
\end{equation}
where $s$ is a spin-index, and 
$\Pi_v (p) \equiv |{\vec \Pi} (p)|  = {\hat p} \cdot {\vec \Pi}(p)$.
Note that $e_N$ and $e_A$ 
are dependent on four momentum $p$ and  are not 
the single particle energies at the on-mass-shell states.

Using the above spinors and energies, we can get the following relations: 
\begin{eqnarray}
(p_0 - e_N(p))(p_0 + e_A(p)) &=& \Pi^2 (p) - M^{*2}(p) ,
\\
 \sum_s u(p,s){\bar u}(p,s) \equiv \Lambda_{+} (p) &=&
\frac{ E_N(p) \gamma_0 -{\vec \Pi}(p)\cdot{\vec \gamma} - M^*(p)}{2E_N(p)} ,
\\
 \sum_s v(-p,s){\bar v}(-p,s) \equiv \Lambda_{-} (p) &=&
\frac{ E_N(p) \gamma_0 +  {\vec \Pi}(p)\cdot{\vec \gamma} + M^*(p)}{2 E_N(p)} .
\end{eqnarray}
Then, the nucleon propagator (\ref{propd}) can be rewritten as
\begin{eqnarray}
G(p) &=&  \sum_{s} ( 1- n(\vp)) 
\frac{u (p,s){\bar u}(p,s)}{p_0 - e_N(p) + i \delta}
 + \sum_{s} n(\vp) 
\frac{u(p,s){\bar u} (p,s)}{p_0 - e_N (p) - i \delta}
\nonumber \\
&& ~~~~~~~~
 + \sum_{s}
\frac{v(-p,s){\bar v}(-p,s)}{p_0 + e_A(p) - i \delta} .
\label{propdN}
\end{eqnarray}
This expression is the same as the single nucleon propagator which was
given  by Bentz et al. as the general form \cite{Bentz}. 
The first, second and third terms exhibit contributions from nucleons 
above the Fermi surface, nucleon in the Fermi sea and negative energy nucleons 
in the Dirac sea, respectively.
Furthermore,
the on-mass-shell positive energy, $\epsi_N$, and 
the negative energy, $\epsi_A$, are defined as the pole energies
of the propagator in Eq.(\ref{propdN}):
\begin{eqnarray}
{\epsi}_N (\vp)&=& e_N (p_0={\epsi}_N ;\vp) ,
\\
{\epsi}_A (\vp)&=& e_A (p_0=-{\epsi}_A ;\vp) .
\end{eqnarray}

We should here note the orthogonal relations between the Dirac spinors.
The Dirac spinors of the positive energy state, $u(p,s)$, and negative energy
state, $v(-p,s)$, are orthogonal only at the same energy, $p_o$:
\begin{equation} 
v^\dagger(-p_0;-\vp,s) u(p_0;\vp,,s^\prime) = 0 .
\end{equation}
but the two spinors on the on-mass-shell condition are not
orthogonal
\begin{equation} 
v^\dagger({\epsi}_A(\vp);-\vp,s) u({\epsi}_N(\vp);\vp,,s^\prime) \neq 0 .
\end{equation}

The positive energy projection operator, $\Lambda_{+} (p)$, 
and the negative energy projection operator, $\Lambda_{-} (p)$,
with the same $p_0$ satisfy the usual relation, 
\begin{equation}
\left\{ \Lambda_{+}(p_0;\vp) + \Lambda_{-}(p_0;\vp)\right\} \gamma_0 = 1 ,
\end{equation}
but those with the on-mass-shell energy do not satisfy this relation: 
\begin{equation}
\left\{ \Lambda_{+}(\epsi_N(\vp);\vp) +
 \Lambda_{-}(-\epsi_A(\vp);\vp)\right\} \gamma_0 \neq 1 .
\end{equation}

This fact tells us a problem in the sum rule.
For an example, the energy non-weighted sum-rule with respect to a
single particle transition operator,  ${\cal O}$, can be obtained with
the single particle wave functions as
\begin{eqnarray}
S &=& \sum_s \int \frac{d^3 p}{(2\pi)^3} n(\vp) 
u^\dagger (\vp,s) {\cal O}^\dagger
{\cal O}u(\vp,s)
\noindent \\
&=&  \sum_s \int \frac{d^3 p}{(2\pi)^3} n(\vp) 
u^\dagger (\vp,s) {\cal O}^\dagger \Lambda_{+} {\cal O} u(\vp,s) + 
 \sum_s \int \frac{d^3 p}{(2\pi)^3} n(\vp) u^\dagger (\vp,s) 
{\cal O}^\dagger \Lambda_{-} {\cal O} u(\vp,s) . 
\label{Nsum}
\end{eqnarray} 
In the RH approximation the first and second terms show the contributions from the
$ph$-states and the {\NNbar}-states, respectively.
In the RHF approximation, however,  
the first term also shows the $ph$-contribution
but the second term dose not completely correspond to the \NNbar-contribution.

\section{Self-Energies in the Relativistic Hartree-Fock approximation}

In this section we should give the detailed expressions of 
the momentum-dependent self-energies.
In order to discuss these effects up to the $N{\bar N}$-threshold energies,
we need to take into account  many kinds of diagrams with multi-meson exchanges.
In this work, however, we aim to demonstrate only qualitative effects of 
the momentum dependence, and then we use the RHF
approximation within the one-meson exchanges.
Furthermore we omit the isovector parts of the self-energies. 
In this work we discuss only qualitative effects of their momentum
dependence in the system with very small asymmetry between the proton
and neutron numbers.

The nucleon self-energy is separated into the Hartree part
and the Fock part as
\begin{equation}
\Sigma(p) = \Sigma_H  + \Sigma_F(p) = - (U_s^H + U_s^F(p))
 +\gamma^\mu ( U_{\mu}^{H} + U_{\mu}^{F} (p), )
\end{equation}
where $\alpha = s, \mu$.
Within the one-boson exchange force, the Fock part of the self-energy
is generally written in the following way \cite{KLW1,RHF,TOMO1}.
\begin{eqnarray}
\Sigma_{F}(p) & = & i \sum_{a} f_a g_a^2 \int \frac{d^{4}k}{(2\pi)^4}
\gamma^{a} G(k) \gamma_{a} {\Delta}_a (p-k) \nonumber \\
& + & i \sum_{b} {\tilde f}_{b} {\tilde g}_b^2 \int \frac{d^{4}k}{(2\pi)^4}
[ (\psla - \ksla), \gamma^{b} ]
G(k) [\gamma_{b}, (\psla - \ksla)] {\Delta}_b (p-k) ,
\label{Fself1}
\end{eqnarray}
where $\gamma_{a(b)}$ is the ${\gamma}$-matrix with the suffix
$a(b)$ indicating the scalar, pseudo-scalar, vector,
axial-vector and tensor,
and $\Delta_a$ is the propagator of meson with the quantum
number indicated with the suffix $a$, written as
\begin{equation}
\Delta_a (q) = \frac{1}{m_a^2 - q^2 - i \delta} .
\end{equation}
In addition $g_a$(${\tilde g}_b$) is a couping constant, and
$f_{a}$(${\tilde f}_b$) is a certain factor including the Fiertz transformation
coefficient in the isospin space and so on.

In general the Fock part of the nucleon self-energy is written as
\begin{eqnarray}
U_{s}^{F} (p) & = &
 \sum_a  C_a^{(s)}
\int \frac{d^3 \vk}{(2 \pi)^3} n({\vk})
\frac{M^* (k)}{ \Pitild_0 (k) } \Delta_a (p-k)   ,
\label{USPV0}
\\
U_{\mu}^{F} (p) & = & U_{\mu}^{(1)} (p)  + U_{\mu}^{(2)} (p)
\end{eqnarray}
with
\begin{eqnarray}
U_\mu^{(1)} (p) & = &
 \sum_a C_a^{(v)} \int \frac{d^3 \vk}{(2 \pi)^3}
n({\vk}) \frac{\Pi_{\mu} (k)}{ \Pitild_0 (k) }
{\Delta}_a (p-k)  ,
\label{UVPV1} \\
U_{\mu}^{(2)} (p) & = &  \sum_b {\tilde C}_b^{(v)}
\int \frac{d^3 \vk}{(2 \pi)^3}
n({\vk}) \frac{\Pi (k) \cdot (p-k)}{ \Pitild_0 (k) } (p_\mu -k_\mu)
{\Delta}_b (p-k)  ,
\label{UVPV2}
\end{eqnarray}
where
\begin{equation}
\Pitild_{\mu} (p) =  \frac{\partial V(p)}{\partial p^{\mu}} .
\end{equation}
Note that some meson exchanges such as the pseudo-scalar meson with the PV
coupling give a different expression which is proportional to 
$(p-k)^2 \Delta(p-k)$.
However these terms can be rewritten to the same expression of the above
equation by introducing the cut-off parameters whose contribution 
can be incorporated into the Hartree parts \cite{tomoGR,tomoDC}. 

According to the way in Ref.\cite{KLW1,TOMO1,tomoGR,tomoDC}, we give 
the Hartree part of the self-energies are given as
\begin{eqnarray}
U_{s}^{H} & = & g_s^H \phi
\\
U_{\mu}^{H} & = & \delta^0_{\mu} \left( \frac{g_v^{H}}{m_v} \right)^2
\rho_H
\end{eqnarray}
where $\phi$ is the scalar mean-field obtained as
\begin{equation}
\frac{\partial}{\partial \phi} {\tilde U} [\phi]
= g^H_s \rho_s
\end{equation}
In the above equations the scalar density $\rho_s$
and the vector Hartree density $\rho_H$ are given by
\begin{eqnarray}
\rho_s & = & 4 \int \frac{{\rm d}^3 {\vp}}{(2 \pi)^3}
n({\vp})
\frac{M_{\alpha}^* (p)}{ \Pitild_0 (p) }   ,
\\
\rho_H & = & 4 \int \frac{{\rm d}^3 {\vp}}{(2 \pi)^3}
n({\vp})
\frac{\Pi_0 (p)}{ \Pitild_0 (p)}   .
\label{rhos}
\end{eqnarray}
The self-energy potential of the $\phi$-field is given
\cite{TOMO1,tomoGR,tomoDC}  by
\begin{equation}
\widetilde{U} [\phi] 
= \frac{\frac{1}{2} m_s^2 \phi^2  + \frac{1}{3} B_s \phi^3
+ \frac{1}{4} C_s \phi^4 }{1 + \frac{1}{2} A_s \phi^2}  .
\eq{sigself}
\end{equation}

Here we comment a relation between the RH
and RHF approximations.
In the RMF approach one usually determines the meson-nucleon coupling
constants to reproduce the saturation properties of the nuclear matter.
When $m_a >> k_F$, the zero range approximation is
available, and  the above Fock terms (\ref{USPV0}) and (\ref{UVPV1}) 
approximately become
\begin{eqnarray}
U_{s}^{F} (p) & = & \sum_a \frac{C_a^{(s)}}{m_a^2}
\int \frac{d^3 \vk}{(2 \pi)^3} n({\vk})
\frac{M^*}{E_k^*} 
=  \left(  \sum_a \frac{C_a^{(s)}}{4 m_a^2} \right) \rho_s ,
\\
U_\mu^{F} (p) & = & \sum_a \frac{C_a^{(v)}}{m_a^2}
 \int \frac{d^3 \vk}{(2 \pi)^3}
n({\vk}) \frac{k^* }{E_k^*}
=  \delta_\mu^0 \left(  \sum_a \frac{C_a^{(v)}}{4 m_a^2} \right) \rho_B .
\end{eqnarray}
These terms are equivalent to the Hartree parts of the self-energies.
Under this approximation, thus, the Hartree and Fock contributions cannot
be distinguished, and 
parameterizing the Hartree parts independently from the Fock parts is the
same as introducing heavy mesons in the RHF calculation.
Therefore we do not need to take the scalar and vector coupling
constants, $g_s^H$ and   $g_v^H$, to be the same ones in the Fock parts.
Indeed Weber et al. \cite{KLW1}  have shown that this method does not break
significant conservation laws as for the nuclear current and the
energy-momentum tensor.

In this work we discuss the self-energies at off-shell energy states, 
where  their imaginary parts appear in some energy region \cite{KLW1}.  
In the actual calculation it is not very easy to calculate the Fock
parts when the imaginary parts appear, and then 
we introduce an additional approximation as follows.

Below the Fermi level with the on-mass-shell condition,
the momentum dependence of the self-energies is not large \cite{RHF}
and can be neglected in actual calculations.
Then we make the self-energies in the integrands in Eqs.(\ref{USPV0}),
(\ref{UVPV1}) and (\ref{UVPV2})  momentum-independent by fixing
their values to be those at the Fermi momentum on the
on-mass-shell condition:  $U_{s(0)} = U_{s(0)}(\epsi_F;k_F)$ and 
$U_i = 0$, where $\epsi_F$ is the Fermi energy;
the similar method was used in the DBHF calculation \cite{DBHF}.
When the self-energies are momentum-independent, 
the tensor-coupling part of the vector self-energy (\ref{UVPV2})
is very small and can be disregarded  \cite{RHF}.
Then the momentum-dependent parts of the self-energies approximately become
\begin{eqnarray}
U_{s}^{F} (p) & = &\int \frac{d^3 \vk}{(2 \pi)^3} n({\vk})
\frac{M^*}{E_k^*}  \sum_a
\frac{ C_a^{(s)}}{m_a^2 - (E_k^* + U_0(k_F) - p_0)^2 + (\vk - \vp)^2 - i \delta} ,
\label{USPA}
\\
U_\mu^{F} (p) & = & \int \frac{d^3 \vk}{(2 \pi)^3}
n({\vk}) \frac{k^* }{E_k^*} \sum_a
\frac{C_a^{(v)}}{m_a^2 - (E_k^* + U_0(k_F) - p_0)^2 + (\vk - \vp)^2 - i \delta} ,
\label{U0PA}
\end{eqnarray}
where $E^*_k = \sqrt{k^2 + M^{*2}(k_F)}$ and $k^*=(E_k^*;\vk)$.

In the actual numerical calculations we consider the Hartree-parts,
$U_s^H$ and $U_0^H$, as fitting parameters, and determine their values
to reproduce the saturation properties, $BE = M - \epsi_N(k_F)$ and
$M^*(k_F)$, at the saturation density.
 
\bigskip

Now we perform the RHF calculation and discuss effects of the
momentum dependence in the self-energies in the off-shell energy region.

When  $\pi$(PV), $\eta$(PV), $\sigma$, $\delta$, $\omega$
meson exchanges are introduced, the Fock terms are described with the
following meson propagations:
\begin{eqnarray} 
\sum_a C_a^{(s)} \Delta_a (q)
&=& -\frac{3}{2}  f_\pi^2 \Delta_\pi -
 \frac{1}{2}f_\eta^2 \Delta_\eta - \frac{1}{2} g_\sigma^2 \Delta_\sigma
 - \frac{3}{2} g_\delta^2 \Delta_\delta 
+  2 g_\omega^2 \Delta_\omega 
+  \frac{12 g_\rho^2 - 9 f_\rho^2}{2} \Delta_\rho 
\\
\sum_a C_a^{(v)} \Delta_a (q)
&=& \frac{3}{2}  f_\pi^2 \Delta_\pi +
 \frac{1}{2}f_\eta^2 \Delta_\eta + \frac{1}{2} g_\sigma^2 \Delta_\sigma
 + \frac{3}{2} g_\delta^2 \Delta_\delta 
+  g_\omega^2 \Delta_\omega 
+  \frac{6 g_\rho^2 + 3 f_\rho^2}{2} \Delta_\rho ,
\end{eqnarray}
where $f_\rho$ is the tensor coupling of the $\rho$-N interaction.
In this work we use the two parameter-sets in the actual numerical 
calculations.

The first set is so called the parameter HF(c), 
which is given in Ref.\cite{RHF}, $g_s^H=g_\sigma=9.116$,
$g_v^H=g_\omega=10.39$, $m_\sigma=550$ MeV, $m_\omega=780$ MeV
and the other couplings being zero.

This  HF(C) parameter-set is standard in the RHF calculation, 
but it includes only the $\sigma$ and $\omega$ mesons.
In the next section, however,  we discuss effects of the Fock terms in
the IFF sum-rule, where all the kinds of mesons contribute to the result. 
Though we do not aim to get any qualitative conclusion in this work,
therefore,
we should use the mass and the couplings of mesons determined
experimentally if possible.
As the second parameter-set, then, we use the couplings and the meson masses 
in the Bonn-A potential \cite{BonnP2} for the Fock parts.
We determine the parameters in the Hartree parts
as $g_s^H = 10.58$,  $g_v^H = 5.338$, $B_s= 26.19$fm$^{-1}$, $C_s=0$ and
$A_s=5.174$fm$^2$ with $m_s = 550$MeV and $m_v=780$MeV.
to reproduce the saturation properties of the nuclear matter 
as the binding energy, $BE=16$ MeV, the incompressibility, $K=250$MeV,
and the effective mass, $M^* = 0.6M$, at the Fermi momentum, 
$k_F = 1.36$fm$^{-1}$.

\begin{figure}
\hspace*{0.5cm}
\includegraphics[scale=0.7]{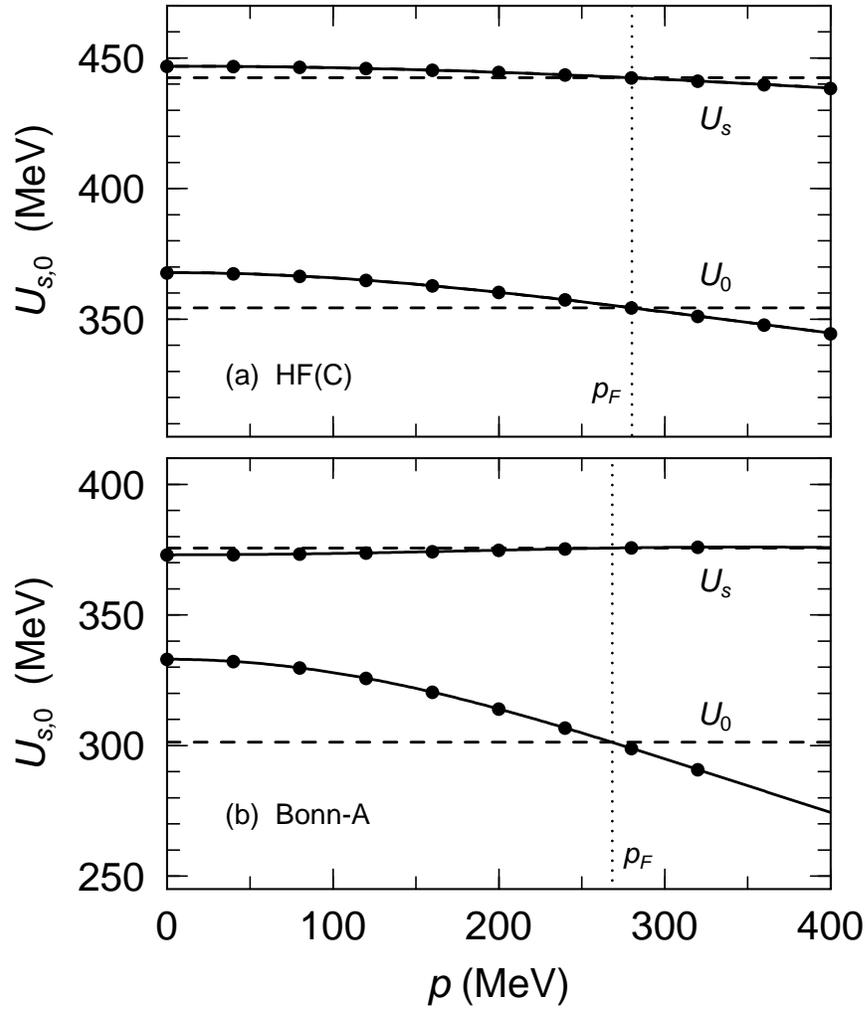}
\caption
{\small 
Momentum-Dependence of the scalar and vector self-energies
with HF(c) and Bonn-A in the upper (a) and lower panels (b).
The solid  and dashed lines represent
the results in RHF and RH, respectively.
The solid circles indicate the results of the full RHF calculation.
The  dotted line   denotes the position of the Fermi momentum.
}
\label{potV}
\end{figure}

We show the momentum dependence of the scalar self-energy $U_s (p)$ and 
that of the time component of the vector self-energy $U_0 (p)$ 
on the on-mass-shell condition with HF(c) in Fig.~\ref{potV}a and with
Bonn-A in Fig.~\ref{potV}b.
The solid and dashed represent the results of our approach, and
the solid circles indicate the results in the fully self-consistent 
RHF calculations with Eqs.(\ref{USPV0}) and (\ref{UVPV1}).
For comparison, we plot $U_s(k_F)$ and $U_0(k_F)$ with the dashed lines; 
we refer to the calculation without the momentum-dependence of the
self-energies as the RH calculation.
In the two parameter-sets, our approximate approach for the Fock terms
in (\ref{USPA}) and (\ref{U0PA}) gives the same results of the full RHF
calculations.  

\begin{figure}[ht]
\hspace*{-1.0cm}
\includegraphics[scale=0.7]{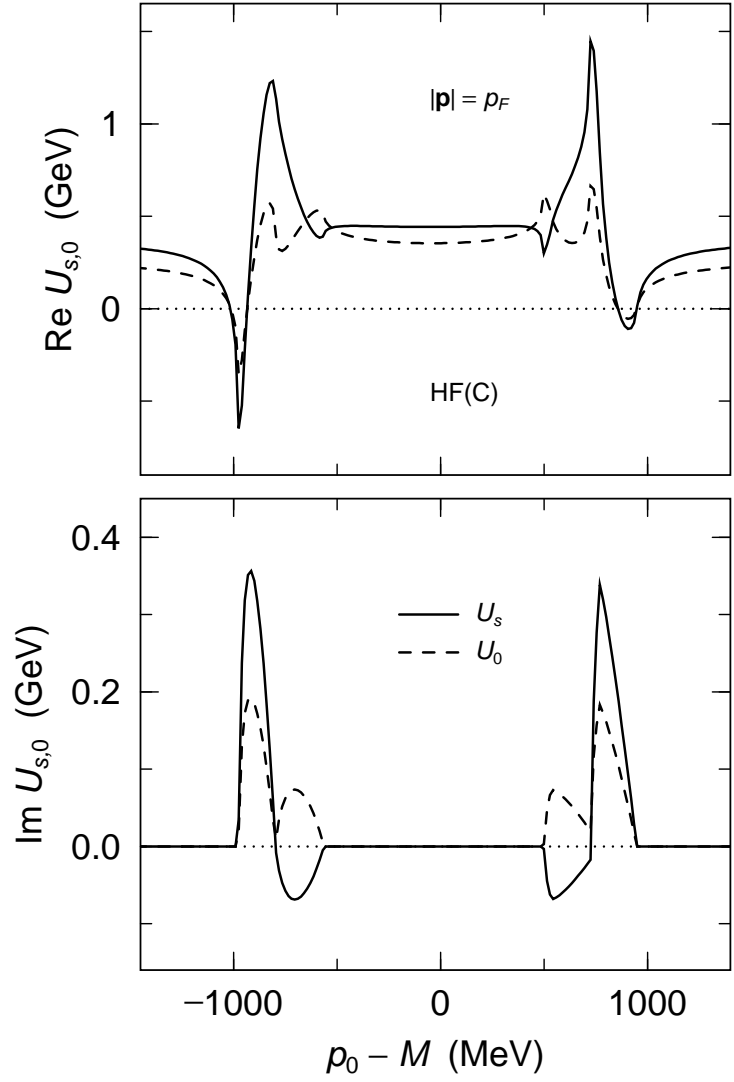}
\caption
{\small 
The scalar and vector self-energies at $|\vp|=k_F$ 
versus energy minus nucleon mass with HF(C).
The solid  and dashed lines represent
the results of the scalar and vector self-energies, respectively.
}
\label{slfEDHF}
\end{figure}

\begin{figure}[ht]
\hspace*{-1.0cm}
\includegraphics[scale=0.7]{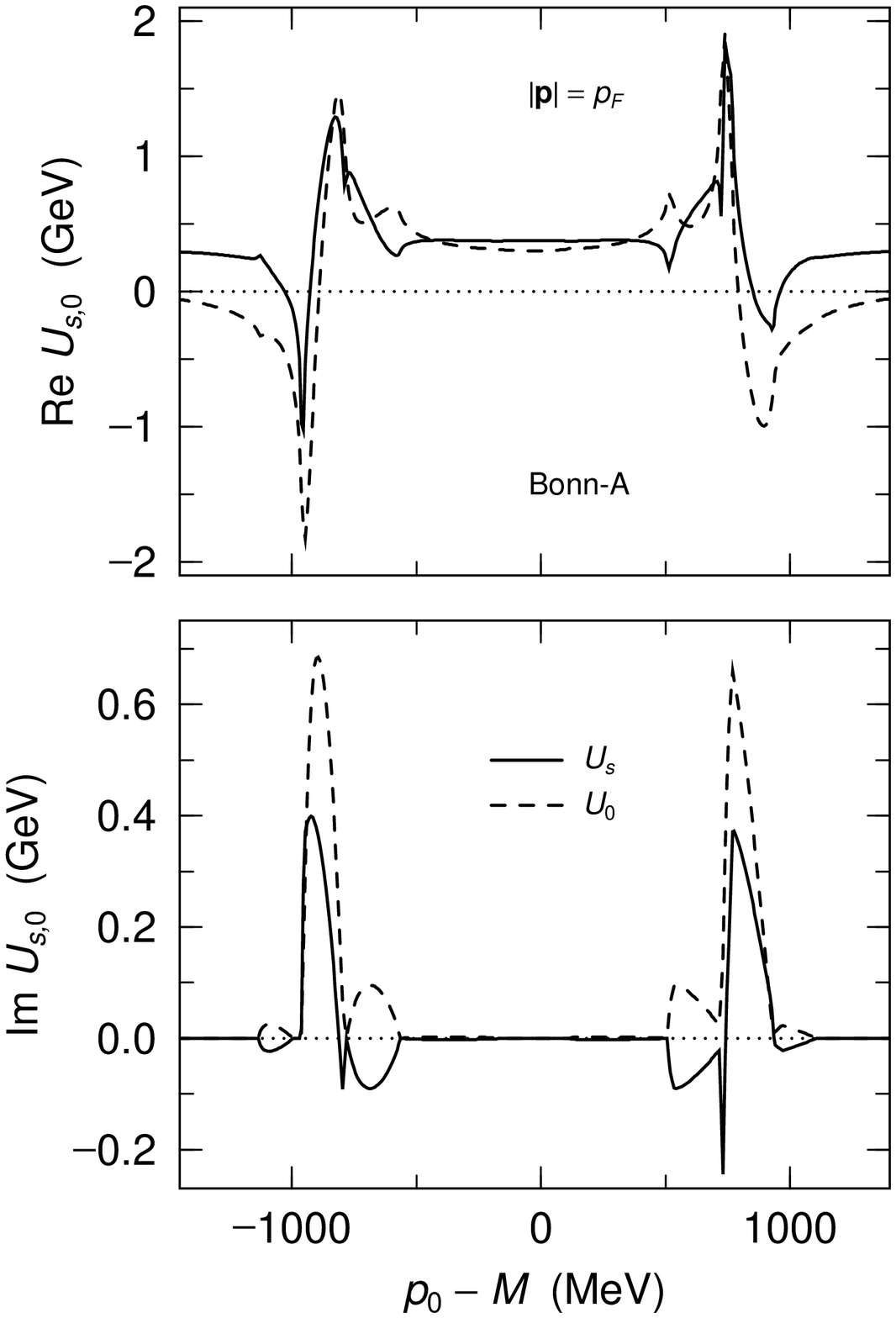}
\caption
{\small 
The same as Fig.~\ref{slfEDHF}, but using the Bonn-A for the parameters.
}
\label{slfEDBn}
\end{figure}

Next we discuss the energy dependence of the self-energies at the fixed
momentum. 
In Fig.~\ref{slfEDHF} we show the scalar self-energies, $U_s(p_0,k_F)$ 
and a time-component of the vector self-energies, $U_0(p_0.k_F)$ 
with HF(C) as functions  of the nucleon energy, $p_0$. 
Furthermore, $U_s(p_0,k_F)$ and  $U_0(p_0.k_F)$ 
with Bonn-A are shown in Fig.~\ref{slfEDBn}.
Their real parts and imaginary parts are plotted 
in the upper and lower panels, respectively.
In the both results with different parameters
the real parts of the self-energies have large energy dependence, 
particularly in the energy region, 
$0.5 {\rm GeV} \gtrsim p_0 - M \gtrsim 1.2 {\rm GeV}$,
where the imaginary parts have large values.
In addition the self-energies around 
the negative energy on-mass shell energy, 
$p_0-M \approx (- 1300)~-~(-1800)$  MeV,
shows the different values from those around the on-mass-shell positive
energy, $p_0 - M \approx - 16$ MeV.

When we calculate observables in low energy phenomena,
on the other hand, the negative energy states, which appear 
as the intermediate states, are in the off-mass-shell energy region.
As an example we consider the situation that off-mass-shell meson with
energy $q_0$ changes to nucleon and antinucleon,
which often appears in one-loop calculation.
When we assume that energy of the nucleon is the Fermi energy $\epsi_F$,
the energy of the antinucleon is at the energy state 
$p_0 =q_0 - \epsi_F$, 
and its self-energies are those on $\epsi_F=q_0$.
In studies about low energy phenomena $|q_0|<<0$, 
the self-energies are considered to be at $p_0 \approx \epsi_F$, and  
they do not have so large momentum-dependence around the Fermi surface.
When we calculate such low energy phenomena, 
hence, we must use the same self-energies of the nucleon 
even for the negative energy nucleon.

Here we give an example.
The energy denominator of the \NNbar-contribution part in
correlation functions is written as
\begin{equation}
D_{\sNNbar} (p,q) = e_N (p) + e_A (p-q) - q_0 .
\end{equation}
When $D_{\sNNbar} << 2M - q_0$,  one considers that 
the \NNbar-production process in the
intermediate state make large contributions to calculational results.

\begin{figure}[ht]
\hspace*{-1.0cm}
\includegraphics[scale=0.8]{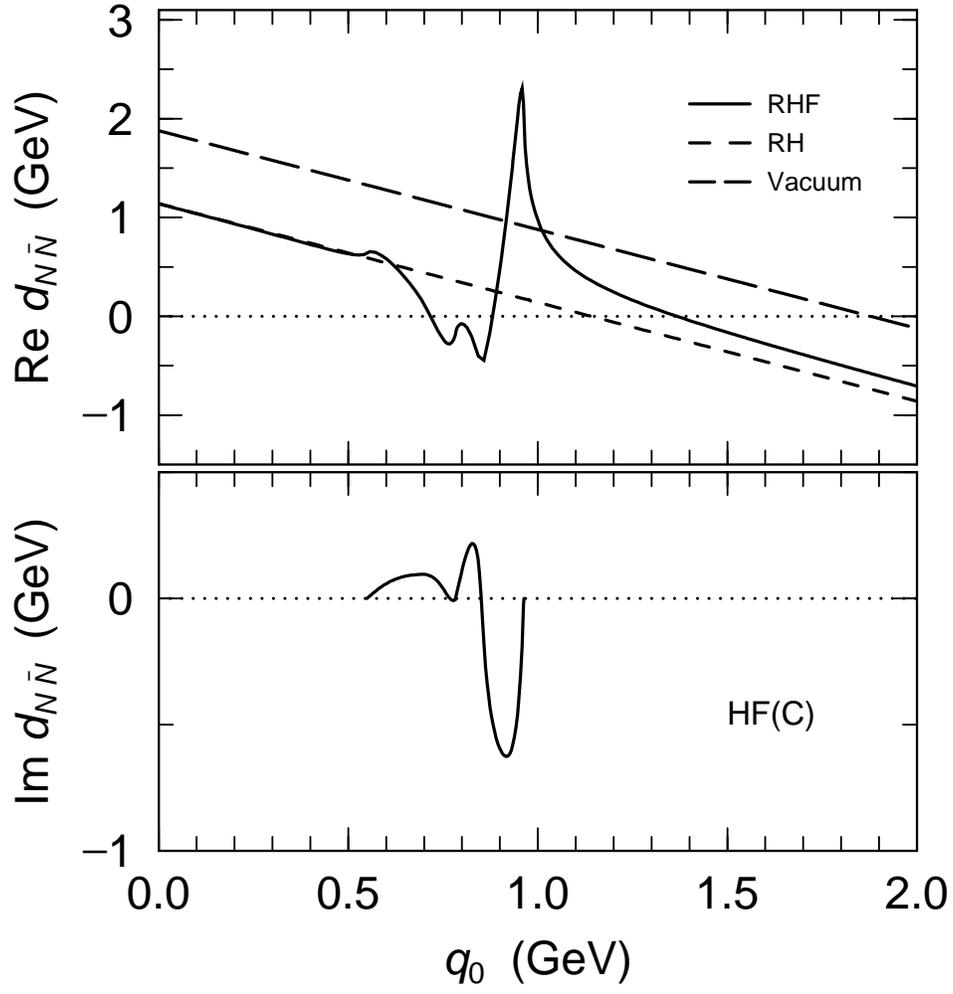}
\caption
{\small 
The real part (a) and the imaginary part (b) of the energy denominator, 
$d_{\sNNbar}$ versus the energy transfer $q_0$ with the paameters, HF(C).
The solid and dashed lines represent the results in the RHF and RH
 approximations, respectively.
The long-dashed line indicate the results in the vacuum.}
\label{EdnHF}
\end{figure}

\begin{figure}[ht]
\hspace*{-1.0cm}
\includegraphics[scale=0.8]{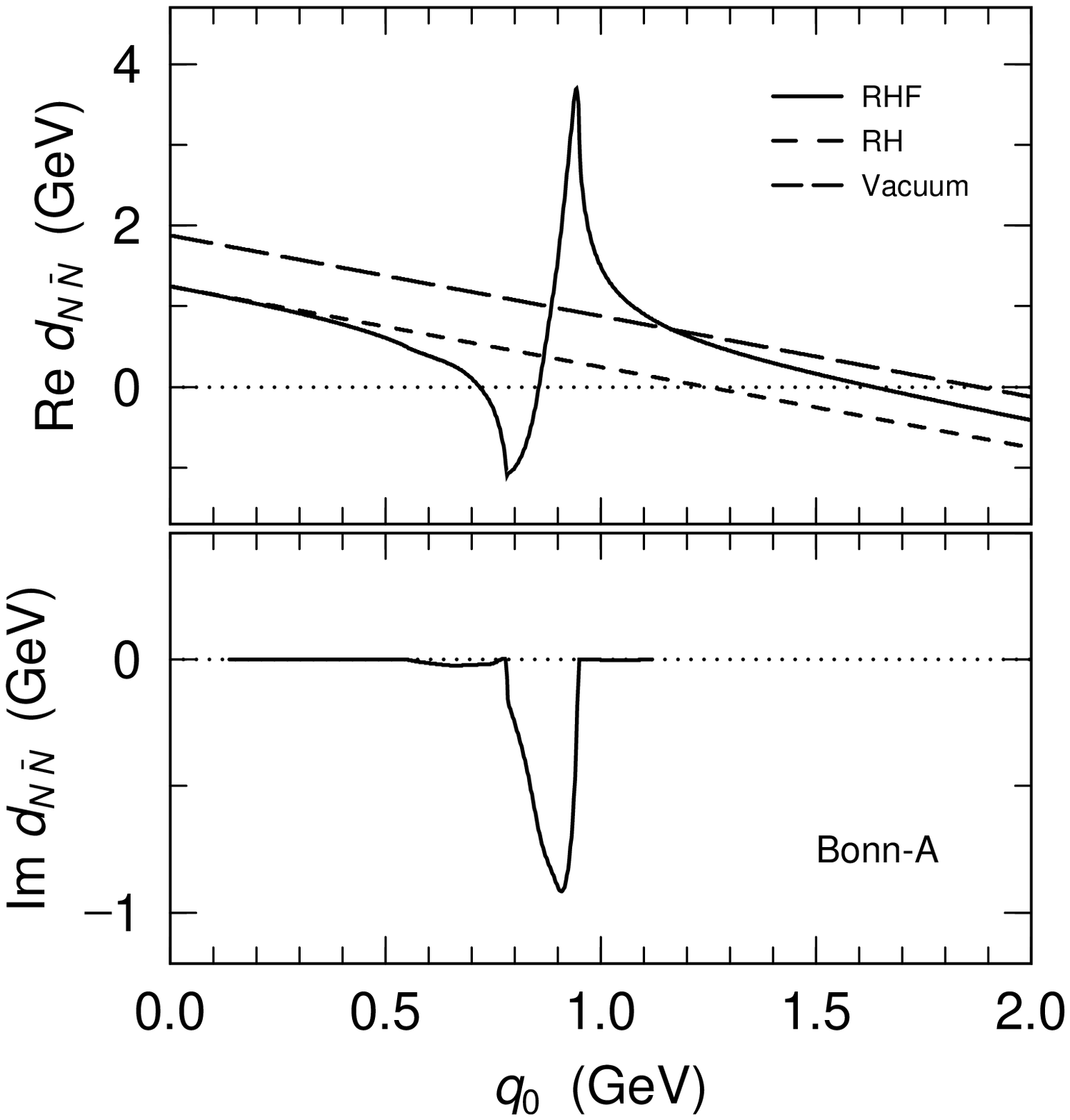}
\caption
{\small 
The same as Fig.~\ref{EdnHF}, but using the Bonn-A for the parameters.
}
\label{EdnBn}
\end{figure}

Here we compare $D_{\sNNbar}(q_0)$ in the RHF and RH
approximations where the self-energies are momentum-independent.
In order to make discussion easy, we calculate them with $\vq = 0$, $|\vp|=k_F$
and $p_0 = e_N = \epsi_F$,
where the above energy denominator becomes 
\begin{equation}
D_{\sNNbar}(\epsi_F;k_F, q_0;0) \equiv 
d_{\sNNbar}(q_0) = \epsi_F + e_A (\epsi_F-q_0, k_F) - q_0.
\end{equation}
We show $d_{\sNNbar}(q_0)$ as a function of the energy transfer, $q_0$, 
with HF(C) in Fig.~\ref{EdnHF} and those with Bonn-A  in Fig.~\ref{EdnBn}. 
The solid and dashed lines represent the results in the RHF and RH
 approximations, respectively.
For comparison we plot  the results in the vacuum, $M^*= M$, with 
the long-dashed line.

In the RH approximation
$d_{\sNNbar}$ has only a real part which is a linear function of $q_0$ and no
imaginary part, while that in the RHF approximation exhibits a different
behavior.
As $q_0$ increases, the real part ${\rm Re} d_{\sNNbar}$ is almost 
the same as that in the RH approximation below $q_0 \approx 0.5$ (GeV),
but shows a quite complicated behavior in $q_0 \gtrsim m_\sigma = 0.55$
GeV, where  the imaginary part becomes large;
the qualitative behaviors are almost same when the parameters are HF(C)
and Bonn-A.

Furthermore, the \NNbar-production threshold energy, 
where $d_{\sNNbar}(q_0) = 0$, is about $q_0 \approx 1250$ MeV in the RH
calculation and $q_0 \approx 1630$ MeV in the RHF calculation.
The threshold energy in the RHF approximation is still smaller 
than that in the vacuum,
$2M \approx 1880$ MeV, but less drastic than that in the RH approximation.

When $q_0 \lesssim 300 MeV$ , the amount of $d_{\sNNbar}$ in the RHF
approximation is almost the same as that in the RH approximation, and
much smaller than that in the vacuum. 
In both the RHF and RH approximations  the threshold energy of the
\NNbar-production is seen to be largely reduced from that in the vacuum
when the energy transfer is very low and far from the on-mass-shell condition.

In high energy transfer region, $q_0 \gtrsim 500$ MeV, however, 
the RH and RHF calculations give quite different results.
The actual threshold energy of the \NNbar-production energy is also
different between two approaches.
Even if the suppressed threshold energy is estimated
from low energy phenomena in the RMF approach, we cannot conclude
that the estimated amount of the threshold energy is the same as that
in the actual \NNbar-production.

\section{Gamow-Teller Strength}

Because of the energy dependence of the nucleon self-energies,
the correlation function must show quite different behaviors 
in the energy region above about 300 MeV 
between the RH and RHF approximations. 
Indeed, Weber et al. \cite{KLW1} suggested that the imaginary part of the
self-energies make another contribution to the nucleon spectral function  
besides the nucleon states on the on-mass-shell positive and negative
energy states.
 
In this section we examine contributions from the Fock terms 
of the self-energies to the GT strength.
Here we calculate the GT strength in the impulse approximation. 
The random phase approximation (RPA) has not given so different amount
of the GT strength from  that in the impulse approximation \cite{ksg1}. 

Here we define the GT-correlation functions as
\begin{equation}
C_{GT}(q) = C_{\beta^{-}} (q) - C_{\beta^{+}} (q)
\label{D-cor}
\end{equation}
with
\begin{eqnarray}
C_{\beta^{-}} (q) & \equiv& C_A (q) = -i \int \frac{d^4 p}{(2 \pi)^4} 
Tr \left\{ G(p+q) \gamma_5 \gamma_2  \tau_{-}  
G(p) \gamma_5 \gamma_2  \tau_{+}  \right\}
\\
C_{\beta^{+}} (q) &=&  -i \int \frac{d^4 p}{(2 \pi)^4} 
Tr \left\{ G(p+q) \gamma_5 \gamma_2  \tau_{+}  
G(p) \gamma_5 \gamma_2  \tau_{-}  \right\} 
\nonumber \\
 &=&  -i \int \frac{d^4 p}{(2 \pi)^4} 
Tr \left\{ G(p-q) \gamma_5 \gamma_2  \tau_{-}  
G(p) \gamma_5 \gamma_2  \tau_{+}  \right\} = C_A(-q).
\label{cor1}
\end{eqnarray}

By substituting Eq.(\ref{propd}) into Eq.(\ref{D-cor}) and disregarding
the vacuum polarization and  the isovector part of the self-energies, 
we can obtain the correlation function as
\begin{eqnarray}
C_{GT} (q) &=&  \frac{4}{(2 \pi)^3} \int \frac{d^3 p}{\Pitild_{0}(p)}
\left[ n^{(n)}(\vp) - n^{(p)}(\vp) \right] \nonumber \\
&& \times
\left[ \frac{t_A(p,q)}{\Pi^2(p+q) - M^{* 2}(p+q) + i \delta}
- \frac{t_A(p,-q)}{\Pi^2(p-q) - M^{* 2}(p-q) + i \delta} \right]
\end{eqnarray}
with
\begin{eqnarray}
t_A (p,q) &=&
\Pi_0(p+q)\cdot \Pi_0(p) - \frac{1}{3}\Pi_v (p+q) \Pi_v(p) + M^{*}(p+q)M^{*}(p)
\end{eqnarray}
where $\Pi_v(p)= \vp \cdot \vPi$.
In addition $n^{(n)}(\vp)  = \Theta(k_n-|\vp|)$ and 
$n^{(p)}(\vp)  = \Theta(k_{p}-|\vp|)$ are the momentum
distributions for neutron and proton, where
 $k_n$ and $k_p$ being the neutron and proton Fermi momenta, respectively.

In the GT-transition we consider only the zero momentum transfer, $\vq = 0$,
and the positive energy transfer, $q_o >0$.
When $|\rho_n - \rho_p| \ll \rho_B$,  in addition, 
we can use the following relation
\begin{equation}
 n^{(n)}(\vp) - n^{(p)}(\vp)  \approx 
\frac{\pi^2}{2 k_F^2} a_r \delta (|\vp| - k_F) 
\end{equation}
with
\begin{equation}
a_r = \frac{2(N - Z)}{A} \rho_B, 
\end{equation}
where $\rho_B$ is the baryon density.
Then the correlation function at $\vq=0$ becomes
\begin{eqnarray}
&& C_{GT} (q_0;\vq=0)
\nonumber \\
&=&
 \frac{a_r}{2 \Pitild_{0}(k_F)}
\left\{
\frac{t_A(p,q)}{E_N(p+q) \left[q_0 + \epsi_F - e_N(p) + i \delta \right]}
\right.
\nonumber \\
&&
~~ -\frac{t_A(p,q)}{E_N(p+q) \left[q_0 + \epsi_F + e_A(p+q) + i\delta \right]}
\nonumber \\
&&~~ - \frac{t_A(p,-q)}{E_N(p-q) \left[q_0 + \epsi_F - e_N(p-q) -i\delta \right]}
\nonumber \\
&&\left.
~~+ \frac{t_A(p,-q)}{E_N(p-q) \left[q_0 + \epsi_F + e_A(p-q) -i \delta \right]}
\right\}_{p = ({\small \epsi_F}; k_F), q=(q_0;0)} .
\end{eqnarray}

The response function is defined as 
\begin{equation}
R_{GT} (q_0) ~=~ - \frac{1}{\pi} {\rm Im} C_{GT}(q_0;\vq=0) .
= a_r \left[ R_{ph} (q_0) + R_{\sNNbar} (q_0) + R_m (q_0) \right]
\label{RespHF}
\end{equation}
with
\begin{eqnarray}
R_{ph} 
&=& \frac{\Pi_0^{2}(k_F) - \frac{2}{3}\Pi_v^{2}(k_F)}{\Pitild_0^{2}(k_F)} 
\delta(q_0)
\\
R_{\sNNbar}& = &
\frac{2\Pi_v^{*2}}{3 \Pitild_0(k_F)\Pitild_0(-\epsi_A(k_F);k_F)}
\delta(q_0 - \epsi_F - \epsi_A(k_F)) .
\end{eqnarray}
In the above equations  $R_{ph}$ and $R_{\sNNbar}$ 
show the contributions from the $ph$-excitation and 
the \NNbar-production, respectively.
The other term, $R_m$, indicates the contribution from the meson production,
which comes from  the imaginary parts of the self-energies.

Using Eq.(\ref{RespHF}) the total GT strength is obtained as
\begin{eqnarray}
S_{tot}& =& \Omega \int_0^{\infty} d q_0 R_{GT} (q_0) 
~=~ S_{ph} + S_{\sNNbar} + S_{meson}
\nonumber \\
&=& 2(N-Z) \left( {\Stil}_{ph} + {\Stil}_{\sNNbar} + {\Stil}_{meson} \right)
\end{eqnarray}
with
\begin{eqnarray}
{\Stil}_{ph} &=& \frac{\Pi_0^{2}(k_F) - \frac{2}{3}\Pi_v^{2}(k_F)}
{\Pitild_0^{2}(k_F)} ,
\\
\Stil_{\sNNbar} &=&
\frac{2\Pi_v^{*2}}{3 \Pitild_0(k_F)\Pitild_0(-\epsi_A(k_F);k_F)} ,
\\
\Stil_{meson} &=& \int_0^{\infty} d q_0 R_{m} (q_0) ,
\end{eqnarray}
where $\Omega$ is the volume of the system.

In the RH approximation the relativistic self-energies are momentum-independent,
and the above response function has been given in Ref.~\cite{ksg1} as 
\begin{eqnarray}
\Stil_{ph} &=& 
1 - \frac{2}{3} \frac{k_F^{2}}{E_F^{*2}} ,
\\
\Stil_{\sNNbar} &=&  \frac{2}{3} \frac{k_F^{2}}{E_F^{*2}} ,
\end{eqnarray}
where the meson production part does not appear, $\Stil_{meson} = 0$.
The $ph$-contribution of the GT-strength is seen to be quenched,
thought
the sum of the two strengths satisfies the IFF sum-rule \cite{ksg1} as
\begin{equation}
S_{ph} + S_{\sNNbar} = S_{IFF} \equiv 2(N-Z) .
\end{equation}

In the RHF framework, on the other hand, the kinetic momentum  $\Pi_\mu$
has energy  dependence even if the spacial component of the momentum
is fixed; 
namely  $\Pi_{0(v)}(\epsi_F;k_F) \neq \Pi_{0(v)}(-\epsi_A(k_F);k_F)$.

When we use the HF(C) parameters, 
$S_{ph}/S_{IFF} \approx 0.84$ and
$S_{ph}/S_{IFF} \approx 0.27$ in the present RHF calculation while
 $S_{ph}/S_{IFF} \approx 0.84$ and
$S_{ph}/S_{IFF} \approx 0.16$ in the RH calculation.
When we use the Bonn-A parameters, 
$S_{ph}/S_{IFF} \approx 0.88 $ and
$S_{ph}/S_{IFF} \approx 0.249$ in the present RHF calculation while
 $S_{ph}/S_{IFF} \approx 0.87$ and
$S_{ph}/S_{IFF} \approx 0.13$ in the RH calculation: 
$ (S_{ph} + S_{\sNNbar})/ S_{IFF}~ \approx 0.11$ 
in the RHF calculation with the above two parameter-sets.

\begin{figure}[ht]
\hspace*{0cm}
\includegraphics[scale=0.8]{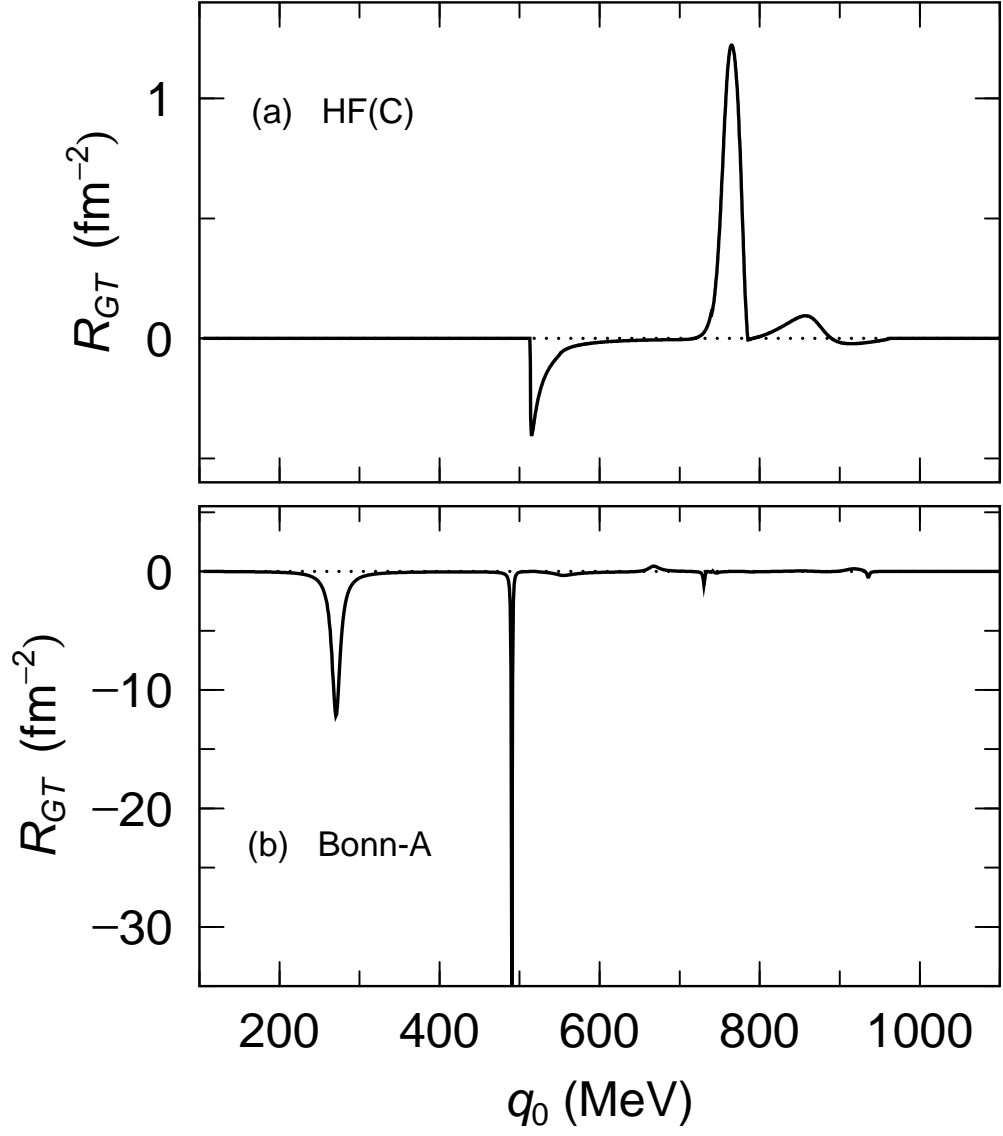}
\caption
{\small 
The GT-response function in the meson production energy region
with HF(C) (a) and Bonn-A (b).
}
\label{RespF}
\end{figure}

As shown before, there is the meson production part of the response function,
$2 a_r R_m(q_0)$,   
in the RHF approximation, which also contributes to the sum-rule.
In Fig.~\ref{RespF} we show the calculational  results of $R_m(q_0)$
in the RHF approximation with HF(C) (a) and  Bonn-A (b).

In this paper we do not intend to discuss the quenching of the
GT-strength, and then 
we do not introduce several important effects such as
the chains of the ring diagrams \cite{ksg2,ks2} appearing in RPA,
diagram of the direct meson production process, 
the delta excitation of nucleon \cite{ss,bentz3} and the two-particle two-hole excitation  \cite{tPtH}.

Thus our results do not give any qualitative conclusions, but
it can successfully demonstrate that 
the suppression of the strength in the $ph$-sector does not
shift only to the \NNbar production
but also to the meson production sector in the relativistic framework.

\section{Summary}

In this paper we calculate the nucleon self-energies
and show their energy dependence in off-mass-shell energy region in the RHF approximation.
We use the two parameter-sets and 
confirm that the calculations with the different parameters give qualitatively 
similar results.

The single particle energies of nucleon and antinucleon appearing 
in the denominator part of the nucleon propagator 
also have energy dependence.
This energy dependence can simultaneously explain the two apparent inconsistent results,
the analysis about the Coulomb sum-rule  \cite{ks1,ks4} and the GT
sum-rule \cite{ksg1,ksg2} and the analysis of the ${\bar p}$-production
experiments \cite{Teis}.  
The former study suggested the large suppression of the
\NNbar-production energy, and the latter one denied it. 
Our result demonstrates that the estimation of the \NNbar-production energy
depends on the energy region.
When we study  low energy phenomena,  
the energy of \NNbar-production in the RHF approximation 
is seen to be almost the same as that in
the the RH approximation and  much smaller
than  that in the vacuum.
In the \NNbar-production, however, the production energy becomes 
larger in the RHF approximation than that in the RH approximation.

Furthermore, we calculate the response function of the GT transition.
In the meson production energy region the nucleon self-energies have
imaginary parts and contribute to the GT response function.
Then 
the quenched strength of the GT transition in the $ph$-excitation sector
does not shift only to the \NNbar-production sector
but also to the meson production sector in the relativistic framework.

For simplicity we consider only contribution from the nucleon
self-energies in the present calculations.
In order to get quantitative conclusions, 
we need to introduce some other effects such as the RPA contributions and
the multi-particle multi-hole excitations and so on.

Furthermore, the RHF calculation with the one-boson-exchange does not give 
the imaginary part of the antinucleon self-energies.
The antinucleon in medium are largely absorbed 
by colliding with nucleons, and its self-energies must have large
imaginary part.
The two contributions from the meson and \NNbar~productions cannot
be either distinguished in actual experiments and more realistic
calculations beyond the Hartree-Fock approximation including multi-boson exchanges.

Anyway the RMF approach based on the RH approximation must be useful to
describe low energy phenomena, and it may be correct that 
the virtual process of the nucleon-antinucleon pair-production plays an
important role there.
However the momentum-dependence of the self-energies must be negligible
below a few hundred MeV at largest, and
we must treat this approach carefully when discussing
phenomena above this energy.


%

\newpage

\end{document}